\documentstyle[art12]{article}
\def\p2t{p_{2\bot}}
\def\q2t{q_{2\bot}}
\begin{document}
\def\emline#1#2#3#4#5#6{%
       \put(#1,#2){\special{em:moveto}}%
       \put(#4,#5){\special{em:lineto}}}
\begin{center}
  \begin{Large}
  \begin{bf}
HEAVY QUARK PHOTOPRODUCTION\\
 IN THE SEMIHARD APPROACH\\
  AT HERA AND BEYOND
  \end{bf}

  \end{Large}
  \vspace{30mm}
  \begin{large}
    V.A.Saleev\\
  \end{large}
Samara State University, Samara 443011, Russia\\
  \vspace{5mm}
  \begin{large}
    N.P.Zotov\\
  \end{large}
Nuclear Physics Institute, Moscow State University,
      Moscow 119899, Russia\\
\end{center}
  \vspace{40mm}
\begin{abstract}
Processes of heavy quark photoproduction at HERA energies and beyond
are investigated using the semihard
($k_{\bot}$ factorization) approach. Virtuality and
longitudinal polarization of gluons in the photon - gluon subprocess
as well as the saturation effects in the gluon distribution
function at small $x$ have been taken into account.
The total cross sections, rapidity and $p_{\bot}$ distributions of
the charm and beauty quark photoproduction have been calculated.
The results are compared with ZEUS experimental data for charm
photoproduction cross section.
\end{abstract}
\newpage

\section{Introduction}

 The heavy quark electro- and photoproduction has become an
increasingly
 important subject of study ~\cite{r1,r2,r3}. In particular
it originates from come into operation of HERA $ep$-collider energy
of which is high
enough to achieve the small $x$ physics region. Indeed
first experimental results obtained at HERA ~\cite{r4} showed
new physical phenomena in  $\gamma $p- and ep-interactions at HERA
energies: resolved photon interactions, jets in photoproduction
produced in
`hard parton collisions, significant fraction of events with a
large
rapidity gap. The measurement of the total
charm
 photoproduction cross section by ZEUS collaboration ~\cite{r5} is
a very important stage of researches.

 The heavy quark production at HERA is a very interesting and important
 subject of study. As it is
dominated by photon-gluon fusion subprocess (Fig.~1) one can
study the gluon distribution functions $G(x, Q^2)$ in small $x$
region (roughly at $x > 10^{-4}$). Secondly this last issue
is important for physics at future colliders (such as LHC): many
processes at these colliders will be determined by small x gluon
distributions.

 At HERA energies and beyond heavy quark production processes
 are of the so-called semihard type ~\cite{r6}. In such
processes by definition a hard scattering scale $Q$ (or heavy quark
mass $M$) is large as compared to the $\Lambda_{QCD}$ parameter,
but $Q$ is much less than the total center of mass energies:
$\Lambda_{QCD} \ll Q \ll \sqrt s$.
Last condition implies that the processes occur in small $x$
region: $x\simeq M^2/s\ll 1$. In such a case the perturbative
QCD expansion has large coefficients $O(\ln^{n}(s/M^2))
\sim O(\ln
^{n}(1/x))$ besides the usual renormalization group ones
which are $O(\ln^{n}(Q^2/\Lambda^2))$.
So, in perturbative QCD the heavy quark photoproduction cross section
(as result of photon-gluon fusion process) has the form ~\cite{r7}:
$\sigma_{\gamma g} = \sigma^{o}_{\gamma g} + \alpha_{s}\sigma^{1}_
{\gamma g}$ + ..., where $\sigma^{1}_{\gamma g}$ was calculated
by Ellis and Nason ~\cite{r8}. The photon - gluon fusion cross
section in low order decreases vs. $s$ at $s\rightarrow\infty:
\sigma^{o}_{\gamma g}\sim M^2/s\ln(s/Q^2)$.
But in the same limit $\sigma^{1}_{\gamma g}\rightarrow Const$,
because heavy quark photoproduction cross section
is dominated by the contribution of the gluon exchange in the
$t$ - channel. These results in breakdown of standard
perturbative QCD expansion and the problem of summing up all
contributions of the order $(\alpha_{s} \ln(Q^2/\Lambda
^2))^{n},(\alpha_{s}\ln(1/x))^{n}$ and $(\alpha_{s}
\ln(Q^2/\Lambda^2)\ln(1/x)^{n}$ in
perturbative QCD appears in calculation of $\sigma_{\gamma g}$.
\par
It is known that summing up the terms of the order $(\alpha_{s}
\ln(Q^2/M^2))^{n}$ in leading logarithm approximation
(LLA) of perturbative QCD leads to the linear DGLAP evolution
equation for deep inelastic structure function ~\cite{r9}.
Resummation of the large contributions of the order of
$(\alpha_{s}\ln(1/x))^{n}$ leads to the BFKL evolution
equation ~\cite{r10} and its solution gives the drastical
increase of the gluon distribution: $xG(x,Q^2)\sim x^{-\omega_{0}},
\omega_{0}=(4N_c\ln2/\pi)\alpha_{s}(Q^2_{0})$.
The sharp growth of the parton density as $x\rightarrow
0$ makes the parton-parton interactions very important,
which in turn makes the so-called GLR evolution equation
 essentially nonlinear  ~\cite{r6}.

The growth of the parton density at $x\rightarrow 0$ and
interactions between partons induce substantial screening
(shadowing) corrections which restore the unitarity constrains
for deep inelastic structure functions (in particular for
a gluon distribution) in small $x$ region ~\cite{r11}.
These facts break the assumption of the standard parton model
(SPM)
about $x$ and transverse momentum factorization for a parton
distribution functions. We should deal with the transverse
momentum factorization ($k_{\bot}$ factorization) theory
{}~\cite{r12,r13}
 or with the so-called semihard approach ~\cite{r11}
 in case of accounting
 for shadowing  corrections at small
$x$ besides taking into account
the virtuality and
the longitudinal polarizations of initial gluons.

In semihard approach ~\cite{r6,r11} screening corrections stop
the growth of the gluon distribution at $x\rightarrow 0$.
This effect was interpreted as saturation of the parton
(gluon) density: gluon distribution function $xG(x,Q^2)$ becomes
proportional to $Q^2R^2$ at $Q^2\leq q^2_{0}(x)$ and $\sigma
\sim \frac{1}{Q^2}xG(x,Q^2)\sim R^2$. The parameters $R$ and
$q^2_{0}(x)$ can be considered as new phenomenological
parameters: the $R$ is related to the size of hadron or of black
spots in a hadron and the parameter $q^2_{0}$ is a typical
transverse momentum of partons in the parton cascade of
a hadron, which leads to natural infrared cutoff in semihard
processes. The parameter $q^2_{0}$ increases with $s$
{}~\cite{r11,r14}. So because of the high parton density in small $x$
region a standard methods of perturbative QCD can not be used
{}~\cite{r6,r11,r14,r15}. In this point the semihard approach
{}~\cite{r6,r11} differs from the works ~\cite{r12,r13,r16} where
after choice of structure function at starting value $Q^2_{0}$ the
QCD evolution is calculated using certain equations.

In the region where the transverse mass of heavy quark $M_{\bot}=
\sqrt{M^2 + p^2_{\bot}}<<\sqrt{s}$ one need to take into
 account the dependence
of the photon-gluon fusion cross section on the virtuality and
polarizations of initial gluon. Thus in semihard approach the
matrix elements of this subprocess differ from the ones of SPM
(see for example ~\cite{r7,r13}).

As for the SPM calculations of the next-to-leading (NLO)
cross section
of the photoproduction of heavy quarks the review
may be found in papers ~\cite{r2,r17}. The results of ~\cite{r8} have been
confirmed by Smith and Van Neerven ~\cite{r18}. The further results
for the electro- and photoproduction of heavy quarks are obtained in
Refs. ~\cite{r19,r20,r21}. Authors of Ref. ~\cite{r17}
note that for beauty quark production the NLO corrections
are large and various estimates of corrections lead to theoretical
uncertainties of the order of a factor of 2 to 3.
As it was estimated
in ~\cite{r22} the total cross section for beauty quark production
at HERA will be only a few tens of percent large than the one-loop
results ~\cite{r8}.

For charm quark theoretical uncertainties are even higher.
It is known that there is also strong mass dependence of
the results of calculations for charm quark production.
Since the mass of the charm quark is small for perturbative
QCD calculations the resumme procedure ~\cite{r12} is less
reliable for charm quark production ~\cite{r22}.

In Refs.~\cite{r23,r24} we used the semihard approach to
calculate the total and differential cross sections of the heavy
quarkonium, $J/\Psi$ and $\Upsilon$, photoproduction. We obtained
the remarkable difference between the predictions of the semihard
approach and the SPM especially for $p_{\bot}$- and $z$-distributions
of the $J/\Psi$ mesons at HERA.

In present paper we investigate the open heavy quark photoproduction
processes in the semihard approach, which was used early in
Ref.~\cite{r11} for calculation of heavy quark production rates
at hadron colliders and for prediction of $J/\Psi$ and$\Upsilon$
photoproduction cross sections at high energies in
Refs.~\cite{r23,r24}.

 Some of the results derived in this paper have already been presented
in Ref. ~\cite{r25}.

\section{QCD Cross Section for Heavy Quark Electroproduction}

\par
 We calculate the total and differential cross sections (the
$p_{\bot}$ and rapidity distributions) of charm and beauty quark
photoproduction
via the photon-gluon fusion QCD subprocess (Fig.1) in the framework
As for the SPM calculations of the next-to-leading (NLO)
cross section
of the semihard ($k_{\bot}$ factorization) approach ~\cite{r6,r11}.
First of all we take into account the transverse momentum of gluon
$\vec{q}_{2\bot}$, its the virtuality $q^2_{2}=-\vec{q}^2_{2\bot}$ and
the alignment of its polarization vectors along its transverse momentum
such as
$\epsilon_{\mu} = q_{2\bot\mu}/\mid\vec{q}_{2\bot}\mid$
{}~\cite{r11,r13}.

Let us define Sudakov variables of the process $ep\to Q\bar Q X$
(Fig.2):
\begin{eqnarray}
p_1&=&\alpha_1 P_1+\beta_1 P_2+p_{1\bot}\qquad
p_2=\alpha_2 P_1+\beta_2 P_2+\p2t\nonumber\\
q_1&=&x_1P_1+q_{1\bot}\qquad
q_2=x_2P_2+\q2t
\end{eqnarray}
where
$$p_1^2=p_2^2=M^2,\qquad q_1^2=q_{1\bot}^2,\qquad q_2^2=\q2t^2,$$
$p_1$ and $p_2$ are 4-momenta of the heavy quarks, $q_1$ is 4-momentum
 of the photon, $q_2$ is 4-momentum of the gluon, $p_{1\bot},~~p_{2\bot},~~
 q_{1\bot},~~q_{2\bot}$ are transverse 4-momenta of these ones.
 In the center of mass frame of colliding particles we can write
 $P_1=(E,0,0,E)$, $P_2=(E,0,0,-E)$, where $E=\sqrt s/2$, $P_1^2=P_2^2=0$
  and $(P_1P_2)=s/2$.
Sudakov variables are expressed  as follows:
\begin{eqnarray}
\alpha_1 &=&\frac{M_{1\bot}}{\sqrt s}\exp(y_1^\ast)\qquad
\alpha_2 =\frac{M_{2\bot}}{\sqrt s}\exp(y_2^\ast)\nonumber\\
\beta_1 &=&\frac{M_{1\bot}}{\sqrt s}\exp(-y_1^\ast)\qquad
\beta_2 =\frac{M_{2\bot}}{\sqrt s}\exp(-y_2^\ast),
\end{eqnarray}
where $M_{1,2\bot}^2=M^2+p_{1,2\bot}^2$, $y_{1,2}^{\ast}$ are rapidities
of heavy quarks, M is heavy quark mass.

 From conservation laws we can easly obtain the following conditions:
\begin{equation}
q_{1\bot}+\q2t=p_{1\bot}+\p2t,\qquad
 x_1=\alpha_1 +\alpha_2,\qquad
 x_2=\beta_1 +\beta_2
\end{equation}
The differential cross section of heavy quark electroproduction has form:
\begin{equation}
\frac{d\sigma}{d^2p_{1\bot}}(ep\to Q\bar Q X)=
\int dy_1^{\ast}dy_2^{\ast}\frac{d^2q_{1\bot}}{\pi}\frac{d^2q_{2\bot}}{\pi}
\frac{|\bar M|^2\Phi_e(x_1,q_{1\bot}^2)\Phi_p(x_2,q_{2\bot}^2)}
 {16\pi^2(x_1x_2s)^2}
 \end{equation}
For photoproduction process it reads:
\begin{equation}
\frac{d\sigma}{d^2p_{1\bot}}(\gamma p\to Q\bar Q X)=
\int dy_1^{\ast}\frac{d^2q_{2\bot}}{\pi}
\frac{\Phi_p(x_2,q_{2\bot}^2)|\bar M|^2}
 {16\pi^2(sx_2)^2\alpha_2}
\end{equation}
We use generalized gluon structure function of a proton
 $\Phi_p(x_2,q_{2\bot}^2)$
which is obtained in semihard approach.  When
integrated over transverse momentum $\vec q_{2\bot}$
$(q_{2\bot}=(0,\vec q_{2\bot},0))$ of gluon up to some limit $Q^2$ it
becomes the  usual structure
function giving the gluon momentum fraction distribution at  scale
$Q^2$:
\begin {equation}
\int\limits_0^{Q^2}\Phi_p
(x,q_{2\bot}^2)d\vec q_{2\bot}^2=xG_p(x,Q^2).
\end{equation}
In our calculation we use the following phenomenological parametrizations
{}~\cite{r11}:
\begin{equation}
\Phi_p(x,q_{2\bot}^2)=C\frac{0.05}{x+0.05}(1-x)^3f_p(x,q_{2\bot}^2),
\end{equation}
\begin{equation}
\Phi_p(x,q_{2\bot}^2)=C_1(1-x)^5f_p(x,q_{2\bot}^2),
\end{equation}
where
\begin{eqnarray}
f_p&=&1,\qquad q_{2\bot}^2\le q_0^2(x)\nonumber\\
f_p&=&(\frac{q_0^2(x)}{q_{2\bot}^2})^2, \qquad q_{2\bot}^2>q_0^2(x),
\end{eqnarray}
and $q^2_{0}(x) = Q^2_{0} + \Lambda^2\exp( 3.56 \sqrt{ \ln(x_0/x)})$,
 $Q_{0}^2 = 2 GeV^2$, $\Lambda = 56$ MeV, $x_{0}$ = 1/3.
The normalization factor $C\simeq 0.97$ mb of the structure function
$\Phi_p(x,q_{2\bot}^2)$ in (7) was obtained in ~\cite{r11} where
$b\bar b$-pair
production at Tevatron energy was described. Thus the normalization
constant C includes the so-called soft $K$--factor,
 which takes into account
$\alpha_s$ corrections due to soft gluon radiation. The typical value
  of $K$--factor in hard hadron-hadron interactions is $K=2-2.5$. As it
  was noted in Ref. \cite{r11} the choice of $C=0.97~$mb corresponds to
  the upper limit of heavy quark hadroproduction cross sections. For the
normalitation factor $C_1$ in (8)
 we shall use value of $C=0.65$ mb, which gives better description
  of existing data on charm quark photoproduction at fixed target
  energies $\sqrt s_{\gamma p}=10-30$ GeV.

 In (4) $\Phi_e(x_1,q_{1\bot}^2)$ is the well known virtual photon spectrum
in Weizsacker-Williams approximation ~\cite{r26} before the integration
over  $ q_{1\bot}^2$:
\begin{equation}
 \Phi(x_1, q_{1\bot}^2)=\frac{\alpha}{2\pi}[\frac{1+(1-x_1)^2}
  {x_1\vec q_{1\bot}^2}-\frac{2m_e^2x_1}{\vec q_{1\bot}^4}].
\end{equation}
In this paper we cosider only heavy quark photoproduction processes.

 The effective gluon distributions $xG(x,Q^2)$ obtained from (7), (8)
increase as $x^{-\omega_0}$  at not very small $x$ $(0.01 < x <
0.15)$,
where $\omega_0 = 0.5$ corresponds to the BFKL Pomeron singularity
{}~\cite{r10}. This rise goes continuously up to $x = x_0$, where
$x_0$ being a solution of the equation $q^2_0(x_0) = Q^2$. In the region
$x < x_0$ there is the saturation of the gluon distribution:
$xG(x,Q^2)\simeq CQ^2$ or $C_1Q^2$.

The square of matrix element of partonic subprocess $\gamma^{\ast} g^{\ast}
\to Q\bar Q$
can be written as follows:
\begin{equation}
|M|^2=16\pi^2e_Q^2\alpha_s\alpha(x_1x_2s)^2[
 \frac{1}{(\hat u-M^2)(\hat t-M^2)}-
\frac{1}{q_{1\bot}^2q_{2\bot}^2}
(1+\frac{\alpha_2\beta_1s}{\hat t-M^2}+
 \frac{\alpha_1\beta_2s}{\hat u-M^2})^2]
\end{equation}
For real photon and off-shell gluon it reads:
\begin{equation}
|M|^2=16\pi^2e_Q^2\alpha_s\alpha(x_2s)^2[\frac{\alpha_1^2+\alpha_2^2}
 {(\hat t-M^2)(\hat u-M^2)}+\frac{2M^2}{q_{2\bot}^2}
(\frac{\alpha_1}{\hat u-M^2}-
 \frac{\alpha_2}{\hat t-M^2})^2],
\end{equation}
where $\alpha_2=1-\alpha_1$ and $\hat s,~\hat t,~~\hat u$ are usual
Mandelstam variables of partonic subprocess
\begin{eqnarray}
 \hat s&=&(p_1+p_2)^2=(q_1+q_2)^2,\qquad
 \hat t=(p_1-q_1)^2=(p_2-q_2)^2,\nonumber\\
 \hat u&=&(p_1-q_2)^2=(p_2-q_1)^2,\qquad
 \hat s+\hat t+\hat u=2M^2+q_{1\bot}^2+q_{2\bot}^2.
\end{eqnarray}

\section{Discussion of the Results}

\par
The results of our calculations for the total cross sections of
$c$- and $b$-quark photoproduction are shown in Figs.~3 and 4.
Solid curves correspond to the semihard approach predictions and
dashed curves correspond to the SPM results with
the GRV LO parametrization of the gluon distribution~\cite{r27}.
We used in our calculations $m_c=1.5$ GeV and $m_b=4.75$ GeV.
Fig.~3a shows us that the solid curve for
charm quark cross section obtained in semihard approach with
 parametrization
(7) describes new ZEUS data ~\cite{r5}
 as well as data from earlier fixed
  target experiments in the range $15 < \sqrt s < 30$ GeV.
The curve of
 the parametrization (8) (Fig.~3b) describes better the
experimental data at low energies ($6 < \sqrt s < 30$ GeV ) but
gives worse description ZEUS data ~\cite{r5}. There are some
 differences between
the theoretical curves and ZEUS experimental data (Figs.~3). If these
ones will
remain valid for new data, then its will show to essential resolved photon
contribution
{}~\cite{r5,r28} and a sensitivity of charm photoproduction cross section
to the photon structure function ~\cite{r29}.

 We would like to note that our results for the parametrization (7)
coincide with the results from Ref. ~\cite{r28}
 for pointlike component of the
charm photoproduction cross section for $MRSD'_{-}$ parton density
at $\mu = m_{c}$ and ones for parametrization (8) correspond to the
charm photoproduction cross section in Ref. ~\cite{r28} for
$MRSA$ parton density.

   At the energy range $\sqrt s_{\gamma p}\ge 500$
GeV the saturation effects in the gluon distribution function
are important for charm quark photoproduction in semihard
approach (Fig.4): the c-quark cross section grows more slowly
at these energies (solid curve). The beauty quark
photoproduction cross section predicted by semihard approach
is larger than the one predicted by SPM at all energies (Fig.~4).
Our conclusion for b-quark photoproduction is the same as in
Ref.~\cite{r11}, where b-quark hadroproduction was described
using semihard approach.

 The $p_{\bot}$ distributions for $c$- and $b$-quark photoproduction
in the semihard approach (solid curves) and in the SPM
(dashed curves) at the energy $\sqrt s_{\gamma p}=200$ GeV are
 shown in Fig.~5.
 The curves are obtained in the semihard
approach for charm quark photoproduction show the saturation effects
in low $p_{\bot}$ region ($p_{\bot} < 2$ Gev/c). At high
$p_{\bot}$ region ($3 < p_{\bot} < 20$ GeV/c the
heavy quark photoproduction $p_{\bot}$ distributions obtained in
the semihard approach are higher than the ones of the SPM (with GRV LO
parametrization of gluon distribution).
This behaviour of c-quark $p_{\bot}$ distributions in the $k_{\bot}$
factorization approach results from the off mass shell
subprocess cross section as well as the saturation
effects of the gluon structure function, because of $m_c^2\approx q_0(x)^2$
at HERA energies. In the case of beauty quarks $m_b^2>>q_0(x)^2$ and
similar effects are absent.

 Fig.6 show the comparison of heavy quark rapidity
distribution (in the photon-proton center of mass frame) obtained
in the different models:
solid curve shows the $y$ distributio
semihard approach,  dashed curve shows the one in SPM. The
effects discussed
above are sufficiently large near the kinematic boundaries,
i.e. at big value of $|y^{\ast}|$. We see that the difference between solid
and dashed curvers can't be degrade at all $y^{\ast}$ via change of
normalization of both models.

\section{Conclusions}

\par
 We showed that the semihard approach describes experimental data
for the charm photoproduction cross section at low and HERA energies,
leads to the saturation effects
for the total cross section of charm quark photoproduction
at $\sqrt s_{\gamma p}\ge 500$ GeV as well as to exceeding over the SPM
prediction
for beauty quark total cross section at the energy range of
$\sqrt s_{\gamma p}=100-500$ GeV, and
predicts a marked difference for rapidity and transverse
momentum distributions of charm and beauty quark photoproduction,
which can be studied already at HERA $ep-$collider.
\vspace{5mm}

{\Large\bf Acknowledgements}

\vspace{4mm}
This  research was supported by the Russian
  Foundation of Basic Research (Grant 93-02-3545).
Authors would like to thank J.Bartels, S.Catani,
G.Ingelman, H.Jung,
J.Lim, M.G.Ryskin and A.P.Martynenko for fruitfull discussions
of the obtained results.

 One of us (N.Z.) would like to thank E.M.Levin
 for discussions of the small $x$
physics and the semihard approach in begin stage of this paper,
L.K.Gladilin and I.A.Korzhavina for discussions of ZEUS
experimental data,
P.F.Ermolov for interest and support
 and also
  W.Buchmuller, G.Ingelman, R.Klanner,
P.Zerwas and the DESY directorate for hospitality and
support at DESY.

                %=======================================================
{\bf Figure captions}
\begin{enumerate}
\item
QCD diagrams for open heavy quark photoproduction subprocesses
\item
Diagram for heavy quark elecrtoproduction
\item
a) The total cross section for open charm quark photoproduction:
solid curve - the semihard approach for the parametrization (7),
dashed curve - the SPM.
The solid dots are the ZEUS measurements \cite{r5} and the open dots
are earlier measurements from fixed target experiments.

b) The solid curve - the semihard approach for the parametrization (8).
  The experimental data as in Fig.~3a
\item
The total cross section for open charm and beauty quark photoproduction:
curves as in Fig.3a

\item
The $p_{\bot}$ distribution for charm and beauty quark photoproduction
at $\sqrt{s_{\gamma p}} = 200$ GeV: curves as in Fig.3a
\item
The $y^{*}$ distribution for charm and beauty quark photoproduction:
at $\sqrt{s_{\gamma p}} = 200$ GeV: curves as in Fig.3a
\end{enumerate}
\newpage

\vspace{30mm}
\unitlength=1.00mm
\special{em:linewidth 1pt}
\linethickness{1pt}
\begin{picture}(129.00,22.00)
\emline{47.00}{7.00}{1}{62.00}{7.00}{2}
\emline{47.00}{7.00}{3}{47.00}{22.00}{4}
\emline{47.00}{22.00}{5}{62.00}{22.00}{6}
\emline{32.00}{22.00}{7}{34.00}{22.00}{8}
\emline{35.00}{22.00}{9}{37.00}{22.00}{10}
\emline{38.00}{22.00}{11}{40.00}{22.00}{12}
\emline{41.00}{22.00}{13}{43.00}{22.00}{14}
\emline{44.00}{22.00}{15}{46.00}{22.00}{16}
\emline{47.00}{7.00}{17}{45.00}{9.00}{18}
\emline{45.00}{9.00}{19}{45.00}{9.00}{20}
\emline{45.00}{9.00}{21}{43.00}{7.00}{22}
\emline{43.00}{7.00}{23}{41.00}{9.00}{24}
\emline{41.00}{9.00}{25}{39.00}{7.00}{26}
\emline{39.00}{7.00}{27}{37.00}{9.00}{28}
\emline{37.00}{9.00}{29}{35.00}{7.00}{30}
\emline{35.00}{7.00}{31}{33.00}{9.00}{32}
\emline{33.00}{9.00}{33}{31.00}{7.00}{34}
\put(26.00,22.00){\makebox(0,0)[cc]{$\gamma$}}
\put(26.00,7.00){\makebox(0,0)[cc]{g}}
\put(69.00,22.00){\makebox(0,0)[cc]{q}}
\put(69.00,7.00){\makebox(0,0)[cc]{$\bar q$}}
\emline{106.00}{7.00}{35}{106.00}{22.00}{36}
\emline{91.00}{22.00}{37}{93.00}{22.00}{38}
\emline{94.00}{22.00}{39}{96.00}{22.00}{40}
\emline{97.00}{22.00}{41}{99.00}{22.00}{42}
\emline{100.00}{22.00}{43}{102.00}{22.00}{44}
\emline{103.00}{22.00}{45}{105.00}{22.00}{46}
\emline{106.00}{7.00}{47}{104.00}{9.00}{48}
\emline{104.00}{9.00}{49}{104.00}{9.00}{50}
\emline{104.00}{9.00}{51}{102.00}{7.00}{52}
\emline{102.00}{7.00}{53}{100.00}{9.00}{54}
\emline{100.00}{9.00}{55}{98.00}{7.00}{56}
\emline{98.00}{7.00}{57}{96.00}{9.00}{58}
\emline{96.00}{9.00}{59}{94.00}{7.00}{60}
\emline{94.00}{7.00}{61}{92.00}{9.00}{62}
\emline{92.00}{9.00}{63}{90.00}{7.00}{64}
\put(85.00,22.00){\makebox(0,0)[cc]{$\gamma$}}
\put(85.00,7.00){\makebox(0,0)[cc]{g}}
\emline{106.00}{7.00}{65}{123.00}{22.00}{66}
\emline{106.00}{22.00}{67}{123.00}{6.00}{68}
\put(129.00,7.00){\makebox(0,0)[cc]{$\bar q$}}
\put(129.00,22.00){\makebox(0,0)[cc]{q}}
\end{picture}

\begin{center}
Fig.~1
\end{center}
\vspace{30mm}

\unitlength=1.00mm
\special{em:linewidth 1pt}
\linethickness{1pt}
\begin{picture}(86.00,40.00)
\emline{41.00}{35.00}{1}{56.00}{35.00}{2}
\emline{56.00}{35.00}{3}{71.00}{40.00}{4}
\emline{41.00}{5.00}{5}{56.00}{5.00}{6}
\put(58.00,5.00){\circle*{5.20}}
\put(64.00,20.00){\circle{8.00}}
\emline{56.00}{35.00}{7}{58.00}{32.00}{8}
\emline{59.00}{31.00}{9}{61.00}{28.00}{10}
\emline{62.00}{27.00}{11}{64.00}{24.00}{12}
\emline{64.00}{16.00}{13}{64.00}{13.00}{14}
\emline{64.00}{13.00}{15}{62.00}{13.00}{16}
\emline{62.00}{13.00}{17}{63.00}{10.00}{18}
\emline{63.00}{10.00}{19}{61.00}{10.00}{20}
\emline{61.00}{10.00}{21}{62.00}{7.00}{22}
\emline{62.00}{7.00}{23}{60.00}{7.00}{24}
\emline{67.00}{23.00}{25}{81.00}{25.00}{26}
\emline{67.00}{17.00}{27}{81.00}{14.00}{28}
\put(33.00,35.00){\makebox(0,0)[cc]{$e(P_1)$}}
\put(33.00,4.00){\makebox(0,0)[cc]{$p(P_2)$}}
\put(54.00,29.00){\makebox(0,0)[cc]{$q_1$}}
\put(54.00,12.00){\makebox(0,0)[cc]{$q_2$}}
\put(86.00,10.00){\makebox(0,0)[cc]{$\bar q(p_2)$}}
\put(86.00,27.00){\makebox(0,0)[cc]{$q(p_1)$}}
\emline{61.00}{5.00}{29}{76.00}{5.00}{30}
\emline{60.00}{3.00}{31}{76.00}{3.00}{32}
\end{picture}

\begin{center}
Fig.~2
\end{center}

\end{document}